\documentclass[11pt]{article}
\usepackage{amssymb}
\usepackage{amsmath}
\usepackage{theorem}
\usepackage{latexsym}
\usepackage{graphicx}
\usepackage[all]{xy}
\reversemarginpar
\theoremheaderfont{\bf} 
\theorembodyfont{\sl}
\newtheorem{theorem}{Theorem}[section]
{\theorembodyfont{\rm} \newtheorem{definition}[theorem]{Definition}}
{\theorembodyfont{\rm} \newtheorem{exa}[theorem]{Example}}
{\theorembodyfont{\rm} \newtheorem{remark}[theorem]{Remark}}
\newtheorem{proposition}[theorem]{Proposition}
\newtheorem{corollary}[theorem]{Corollary}

{\theorembodyfont{\rm} \newtheorem{ass}[theorem]{General Assumption}}

{\theorembodyfont{\rm}}
{\theorembodyfont{\rm}}
\newenvironment{proof}{{\sc Proof:}}{\mbox{}\hfill$\Box$\par}
\newcommand{\eqnref}[1]{~\mbox{$(${\rm \ref{#1}}$)$}}

\renewcommand{\theequation}{\thesection.\arabic{equation}}

\newcommand{\TS}{\textstyle}
\newcommand{\N}{{\mathbb N}}

\newcommand{\F}{{\mathbb F}}
\newcommand{\C}{{\mathbb C}}

\newcommand{\cC}{{\mathcal C}}

\newcommand{\cF}{{\mathcal F}}
\newcommand{\cG}{{\mathcal G}}

\newcommand{\cM}{{\mathcal M}}

\newcommand{\cP}{{\mathcal P}}
\newcommand{\cH}{{\mathcal H}}

\newcommand{\HH}{\mbox{\rm\textbf H}}
\newcommand{\cCconst}{\mbox{${\mathcal C}_{\,\text{\rm const}}$}}
\newcommand{\cCperpconst}{\mbox{$\widehat{\mathcal C}_{\,\text{\rm const}}$}}
\newcommand{\cCcoeff}{\mbox{${\mathcal C}_{\,\text{\rm coeff}}$}}
\newcommand{\cCperpcoeff}{\mbox{$\widehat{\mathcal C}_{\,\text{\rm coeff}}$}}
\newcommand{\rank}{\mbox{\rm rank}\,}
\newcommand{\edge}[2]{\mbox{$-\!\!\!-\!\!\!\longrightarrow$}\hspace{-1.8em}%
\raisebox{1.6ex}{${\scriptscriptstyle (\!\frac{\,{#1}\,}{#2}\!)}$}\hspace{.8em}}
\newcommand{\im}{\mbox{\rm im}\,}

\newcommand{\wt}{\mbox{\rm wt}}
\newcommand{\we}{\mbox{\rm we}}

\newcommand{\rev}{\mbox{\rm rev}}

\newcommand{\ddual}[1]{\mbox{$\langle\langle{#1}\rangle\rangle$}}
\newcommand{\T}{\mbox{$^{\sf T}$}}

\newcommand{\Smallfourmat}[4]{\mbox{\scriptsize{$\begin{pmatrix}{#1}&{\!\!\!#2}\\{#3}&{\!\!\!#4}
                     \end{pmatrix}$}}}

\newcounter{abc}

\newenvironment{romanlist}{\begin{list}{{\rm (\roman{abc})}\hfill}{\usecounter{abc}
     \topsep0ex \labelwidth.7cm \leftmargin.7cm \labelsep0cm
     \rightmargin0cm \parsep0ex \itemsep.6ex
     \partopsep1.6ex}}{\end{list}}
\newenvironment{alphalist}{\begin{list}{{\rm (\alph{abc})}\hfill}{\usecounter{abc}
     \topsep0ex \labelwidth.6cm \leftmargin.6cm \labelsep0cm
     \rightmargin0cm \parsep0ex \itemsep.6ex
     \partopsep1.6ex}}{\end{list}}

\title{A MacWilliams Identity for Convolutional Codes: \\[.7ex]
              The General Case}
\date{May 22, 2008}
\author{Heide Gluesing-Luerssen\footnote{University of Kentucky, Department of Mathematics, 715 Patterson
        Office Tower, Lexington, KY 40506-0027, USA; heidegl@ms.uky.edu},
        Gert Schneider\footnote{University of Groningen, Department of Mathematics, P.~O.~Box 407,
       9700 AK Groningen, The Netherlands; schneider@math.rug.nl}
       }

\begin{document}
\maketitle
\noindent{\bf Abstract:}
A MacWilliams Identity for convolutional codes will be established.
It makes use of the weight adjacency matrices of the code and its dual,
based on state space realizations (the controller canonical form) of the codes in question.
The MacWilliams Identity applies to various notions of duality appearing
in the literature on convolutional coding theory.



\section{Introduction}\label{S-Intro}
\setcounter{equation}{0}
The weight enumerator of a code keeps track of the distribution of codeword weights and therefore is
of paramount importance for the error-correcting capabilities of the code in question.
One of the most celebrated results in block code theory, the MacWilliams Identity Theorem, states that
the weight enumerator of a block code completely determines the weight enumerator of the dual code and gives an
explicit transformation formula.
The practical and theoretical implications for block code theory have been studied ever since, see for instance
\cite[Ch.~11.3, Ch.~6.5, Ch.~19.2]{MS77} or \cite[Thm.~7.9.5]{HP03}.

For convolutional codes only partial results concerning a possible MacWilliams Identity could be established so far.
About~$30$ years ago it has been shown by a simple example that the classical weight enumerator as introduced
by Viterbi~\cite{Vi71} does not obey any MacWilliams type of identity, see~\cite{SM77}.
In other words, this weight enumerator is too coarse in order to yield detailed information about the dual code.
This insight gave rise to the study of a more refined weight enumerating object, the weight adjacency matrix
(WAM).
It has been introduced in~\cite{McE98a}, but appears already in different notations earlier in the literature.
Indeed, one can show that it basically coincides with the labels of the weight enumerator state diagram as
considered in~\cite{Ab92}.
The WAM is defined via a state space description of the encoder as introduced in~\cite{MaSa67}.
It is labeled by the set of all state pairs $(X,Y)$, and each entry contains the weight enumerator of all
outputs associated with the corresponding state transitions from~$X$ to~$Y$.
The resulting matrix contains considerably more information about the code than the classical weight enumerator
mentioned above.
Indeed, it is well-known \cite{McE98a},~\cite{GL05p} how to derive the latter from the WAM.
Unfortunately, the matrix by itself is not an invariant of the code, but rather depends on the choice of the
encoder and the state space realization.
However, this dependence can nicely be described and upon factoring out a suitable group action results in an
invariant of the code, the generalized WAM.

In a previous article~\cite{GS08} we studied this invariant in detail, and, in particular, we could
establish a weak MacWilliams type of identity for the generalized WAM.
It states that a certain transform of any WAM of a given code results in a matrix having
up to ordering the same entries as any WAM of the dual code.
For the class of codes with all Forney indices being at most~$1$ we could even show that this ordering is
actually induced by a state space isomorphism, which can also be given explicitly.
Of course, the isomorphism depends on the chosen representations of the code and its dual.
This result generalizes a MacWilliams identity established in~\cite{Ab92} for the class of codes of degree~$1$ (that is,
only one Forney index has the value~$1$ while all other indices are zero).

In this paper we will extend the result to arbitrary CC's.
In other words, we will establish a MacWilliams Identity for the full class of CC's.
Stated more precisely, given a code and its dual with chosen state space representations we will give an
explicit transformation of the WAM that will result in the WAM of the dual code.
The result generalizes the classical MacWilliams Identity for block codes.

The main outline of the paper is as follows.
In the next section we will introduce the basic notions of convolutional coding theory including
state space realizations as well as two block codes closely related to the given CC.
In Section~\ref{S-adjmatrix} we will introduce the WAM as well as the
MacWilliams transformation matrices, and we will state the MacWilliams Identity.
Section~\ref{S-Proof} will be completely devoted to the proof of the MacWilliams Identity and therefore will
be rather technical.
A detailed example will illustrate the steps of the MacWilliams transformation.
Finally, in Section~\ref{S-sequence} we will discuss an alternative notion of duality for CC's
and translate our result to that notion.

The following notation will be used throughout.
For any domain~$R$ and any matrix $M\in R^{a\times b}$ we denote by $\im M:=\{uM\mid u\in R^a\}$ and
$\ker M:=\{u\in R^a\mid uM=0\}$
the image and kernel, respectively, of the canonical linear mapping $R^a\longmapsto R^b,\ u\longmapsto uM$
associated with~$M$.

\section{Preliminaries}\label{S-Prelim}
\setcounter{equation}{0}
In this section we will collect the main notions of convolutional coding theory as needed for this paper.
Let~$\F$ be a finite field.
A {\em $k$-dimensional convolutional code of length\/} $n$ is a submodule~$\cC$ of $\F[D]^n$ of the form
\[
    \cC=\im G:=\{uG\,\big|\, u\in\F[D]^k\}
\]
where~$G$ is a {\em basic\/} matrix in $\F[D]^{k\times n}$, i.~e., there exists a matrix
$\tilde{G}\in\F[D]^{n\times k}$ such that $G\tilde{G}=I_k$.
In other words,~$G$ is noncatastrophic and delay-free.
We call $G$ an {\em encoder\/} and the number
$\delta:=\max\{\deg\gamma\mid\gamma\text{ is a $k$-minor of }G\}$ is said to be the
{\em degree\/} of the code~$\cC$.
A code having these parameters is called an $(n,k,\delta)$ code.
A basic matrix $G\in\F[D]^{k\times n}$ with rows $g_1,\ldots, g_k\in\F[D]^n$ is said to be
{\em minimal\/} if $\sum_{i=1}^k\deg (g_i)=\delta$.
For characterizations of minimality see, e.~g., \cite[Main~Thm.]{Fo75} or \cite[Thm.~A.2]{McE98}.
It is well-known~\cite[p.~495]{Fo75} that each code~$\cC$ admits a minimal encoder~$G$.
The row degrees $\deg g_i$ of a minimal encoder~$G$ are uniquely determined up to ordering and are called the
{\em Forney indices\/} of the code or of the encoder.
It follows that a CC has a constant encoder matrix if and only if the degree is zero.
In that case the code can be regarded as a block code.

Throughout the main part of this paper the dual of a code $\cC\subseteq\F[D]^n$ is defined as
\begin{equation}\label{e-Cperpdef}
  \widehat\cC:=\{w\in\F[D]^n\mid wv^{\sf T}=0\text{ for all }v\in\cC\}.
\end{equation}
In other words, the dual code~$\widehat{\cC}$ is the orthogonal of~$\cC$ with respect to the $\F[D]$-bilinear form
$\big((w_1,\ldots,w_n),\,(v_1,\ldots,v_n)\big)\longmapsto \sum_{i=1}^nw_iv_i\in\F[D]$.

In Section~\ref{S-sequence} we will address a different notion of duality that has been
introduced in the literature on CC's as well, and we will show how our
result can be translated to that notion.
Duality as defined in\eqnref{e-Cperpdef} has been considered in, e.~g.,
\cite{Ab92}, \cite{Fo73}, \cite{Fo91},~\cite{McE98}, and ~\cite{SM77}.
It is well known \cite[Thm.~7.1]{McE98} that
\begin{equation}\label{e-Cperp}
 \text{if $\cC$ is an $(n,k,\delta)$ code, then $\widehat\cC$ is an
       $(n,n-k,\delta)$ code.}
\end{equation}

For a block code $V\subseteq\F^\ell$ the dual is denoted by
$V^{\perp}:=\{w\in\F^\ell\mid wv^{\sf T}\!=0\;\text{for all}\; v\in V\}$.
The different notation $\widehat{\cC}$ versus $V^{\perp}$ for the dual of a convolutional code~$\cC$ versus
the dual of a block code~$V$ will be helpful later on.

The weight of convolutional codewords is defined straightforwardly.
For a polynomial vector $v=\sum_{j=0}^Nv^{(j)}D^j\in\F[D]^n$ we define
$\wt(v):=\sum_{j=0}^N\wt(v^{(j)})$, where $\wt(v^{(j)})$ is the Hamming weight
of the constant vector $v^{(j)}\in\F^n$.
Let $\C[W]_{\leq n}$ denote the vector space of polynomials over~$\C$ in the indeterminate~$W$
of degree at most~$n$.
For any subset $S\subseteq\F^n$ we define the {\em weight enumerator\/} of~$S$
to be the polynomial
\[
    \we(S):=\sum_{j=0}^n \alpha_jW^j\in\C[W]_{\leq n}, \text{ where }\alpha_j:=\#\{a\in S\mid \wt(a)=j\}.
\]
Recall that the classical MacWilliams Identity for block codes states that
if $\cC\subseteq\F^n$ is a $k$-dimensional code and $\F=\F_q$ is a field with~$q$ elements, then
\begin{equation}\label{e-MacWBC}
   \we(\cC^{\perp})=q^{-k}\HH\big(\we(\cC)\big)
\end{equation}
with $\HH$ being the MacWilliams transform
\begin{equation}\label{e-h}
  \HH:\, \C[W]_{\leq n}\longrightarrow \C[W]_{\leq n},\quad
       \HH(f)(W):=(1+(q-1)W)^n f\big({\TS\frac{1-W}{1+(q-1)W}}\big).
\end{equation}
It should be kept in mind that~$\HH$ depends on the parameters~$n$ and~$q$.
Since throughout this paper these parameters will be fixed we do not indicate them explicitly.

A central tool for the purpose of our paper is the description of a CC by
the controller canonical form (CCF).
It will allow us to introduce the main object, the WAM,
as well as two block codes associated with a CC that are crucial for our investigation.
Even though the CCF can be found in any textbook on control theory, we choose to present it here
explicitly since many of our matrix identities later on will rely on the precise form of the matrices.

\begin{definition}\label{D-CCF}
Let $G\in\F[D]^{k\times n}$ be a minimal encoder with Forney indices
$\delta_1,\,\ldots,\delta_r>0=\delta_{r+1}=\ldots=\delta_k$
and degree $\delta:=\sum_{i=1}^k\delta_i$.
Let~$G$ have the rows $g_i=\sum_{\nu=0}^{\delta_i}g_{i,\nu}D^{\nu},\,i=1,\ldots,k,$ where
$g_{i,\nu}\in\F^n$.
For $i=1,\ldots,r$ define the matrices
\[
 A_i=\left(\begin{smallmatrix} 0&1& & \\ & &\ddots& \\& & &1\\ & & &0\end{smallmatrix}\right)
      \in\F^{\delta_i\times\delta_i},\
 B_i=\begin{pmatrix}1&0&\cdots&0\end{pmatrix}\in\F^{\delta_i},\
 C_i=\begin{pmatrix}g_{i,1}\\ \vdots\\ g_{i,\delta_i}\end{pmatrix}\in\F^{\delta_i\times n}.
\]
The {\em controller canonical form (CCF)\/} of~$G$ is defined as
the matrix quadruple
$(A,B,C,E)\in\F^{\delta\times\delta}\times\F^{k\times\delta}\times
             \F^{\delta\times n}\times\F^{k\times n}$
where
\[
   A=\left(\begin{smallmatrix} A_1&  & \\ &\ddots &\\ & &A_r\end{smallmatrix}\right),\:
   B=\begin{pmatrix}\bar{B}\\0\end{pmatrix}\text{ with }
   \bar{B}=\left(\begin{smallmatrix}
            B_1\!\!& &\\ &\ddots & \\ & &\!\!B_r\end{smallmatrix}\right),\:
   C=\left(\begin{smallmatrix}C_1\\ \vdots\\C_r\end{smallmatrix}\right),\:
   E=\left(\begin{smallmatrix}g_{1,0}\\ \vdots\\g_{k,0}\end{smallmatrix}\right)=G(0).
\]
We call $(A,B,C,E)$ a CCF of the code $\cC\subseteq\F[D]^n$ if $(A,B,C,E)$ is the
CCF of a minimal encoder of~$\cC$.
\end{definition}

It is well-known that the CCF describes the encoding process of the matrix~$G$ in form
of a state space system.
Indeed, $G(D)=B(D^{-1}I-A)^{-1}C+E$, see~\cite[Prop.~2.1, Thm.~2.3]{GL05p}.
As a consequence, one has for $u=\sum_{t\geq 0}u_tD^t\in\F[D]^k$ and $v=\sum_{t\geq0}v_tD^t\in\F[D]^n$
\begin{equation}\label{e-SSS}
   v=uG
  \Longleftrightarrow
  \left\{\begin{array}{rcl} x_{t+1}&=&x_tA+u_tB\\v_t&=&x_tC+u_tE\end{array}
    \;\text{ for all }t\geq0\right\} \text{ where }x_0=0.
\end{equation}
We call~$\F^{\delta}$ the {\em state space\/} of the encoder~$G$ (or of the CCF) and $x_t\in\F^{\delta}$
the {\em state at time\/}~$t$.

The following two block codes are naturally associated with a given code.

\begin{definition}\label{D-Cconst}
For a code $\cC\subseteq\F[D]^n$ define the associated block codes
$\cCconst:=\cC\cap\F^n$ and
$\cCcoeff:=\big\{w\in\F^n\,\big|\,\exists\; v=\sum_{t\geq 0}v_tD^t\in\cC\;\text{such that}\;
         v_{\hat{t}}=w\;\text{for some}\;\hat{t}\geq0\big\}$.
\end{definition}

Obviously, $\cCconst$ is simply the block code consisting of the constant codewords in~$\cC$.
Consequently, this space is generated by the constant rows (if any) of a minimal encoder matrix~$G$.
The code~$\cCcoeff$ is the space of all constant vectors that appear as coefficient vectors of some
codeword.
It can easily be described by using a CCF $(A,B,C,E)$ for~$\cC$.
Indeed, let $G=\sum_{t\geq0}^mG_tD^t$, where $G_t\in\F^{k\times n}$.
Then obviously $\cCcoeff=\im (G_0^{\sf T},G_1^{\sf T},\ldots,G_m^{\sf T})^{\sf T}$.
Since the coefficient vectors of the rows of~$G$ are collected in the matrices $C$ and~$E$, this yields
\begin{equation}\label{e-CC}
   \cCcoeff=\im\begin{pmatrix}C\\E\end{pmatrix}.
\end{equation}


In \cite[Prop.~II.7]{GS08} it has been shown that the two block codes from Definition~\ref{D-Cconst}
and the corresponding codes $\cCperpcoeff$ and $\cCperpconst$
associated with the dual code~$\widehat{\cC}$ are crosswise mutual duals.
Precisely, we have the following result.

\begin{proposition} \label{P-duality}
Let $\cC$ be an $(n,k,\delta)$ code over~$\F_q$ with $r$ positive Forney indices, and let the
dual $\widehat{\cC}$ have $\widehat{r}$ positive Forney indices.
Then $\dim\cCconst=k-r$, $\dim\cCperpconst=n-k-\widehat{r}$, and $\dim \cCcoeff=k+\widehat{r}$,
$\dim\cCperpcoeff=n-k+r$.
Furthermore, $(\cCcoeff)^\perp=\cCperpconst$, and, consequently,
$q^{k+\hat{r}}\we(\cCperpconst)=\HH\big(\we(\cCcoeff)\big)$.
\end{proposition}

\section{The Weight Adjacency Matrix of a Code}\label{S-adjmatrix}
\setcounter{equation}{0}
The weight adjacency matrix as defined below has been introduced in \cite{McE98a} and
studied in detail in \cite{GL05p} as well as~\cite{GS08}.
The aim of this section is to present the basic properties of the weight adjacency
matrix for a given CC as well as to formulate our main result.

Recall from\eqnref{e-SSS} that the controller canonical form of an encoder leads to a state space
description of the encoding process where the input is given by the coefficients of the message stream
while the output is the sequence of codeword coefficients.
The following matrix collects for each possible pair of states~$(X,Y)$ the information whether via
a suitable input~$u$ a transition from~$X$ to~$Y$ is possible, i.~e., whether $Y=XA+uB$ for some~$u$,
and if so, collects the weights of all associated outputs $v=XC+uE$.

\begin{definition}\label{D-Lambda}
Let $G\in\F[D]^{k\times n}$ be a minimal encoder with CCF $(A,B,C,E)$.
The {\em weight adjacency matrix (WAM)\/}
$\Lambda:=\Lambda(G)\in\C[W]^{q^{\delta}\times q^{\delta}}$ of~$G$ is
defined to be the matrix indexed by $(X,Y)\in\F^{\delta}\times\F^{\delta}$ with the entries
\begin{equation}\label{e-LambdaXY}
     \Lambda_{X,Y}:=\we(\{XC+uE\mid u\in\F^k: Y=XA+uB\})\in\C[W]_{\leq n}.
\end{equation}
\end{definition}
Observe that if $\delta=0$ the matrices $A,\,B,\,C$ do not exist while $E=G$.
As a consequence, $\Lambda=\Lambda_{0,0}=\we(\cC)$ is the ordinary weight enumerator of the block code
$\cC=\{uG\mid u\in\F^k\}\subseteq\F^n$.

The WAM contains very detailed information about the code.
The classical path weight enumerator\cite[p.~154]{JoZi99}, the extended row distances~\cite{JPB90},
the active burst distances~\cite{HJZ02} as well as the column distances of the code can all be
computed from the WAM, see~\cite{McE98a},~\cite[Sec.~3.10]{JoZi99} and~\cite{GL05p}.
For the relevance of these distance parameters for the error-correcting performance of the code
see~\cite{JPB90},~\cite{HJZ02}

Let us illustrate the matrix by an example.

\begin{exa}\label{E-1}
Let $\F=\F_3$ and
\[
    G=\begin{pmatrix}1+D^2&2+D&0\\1&0&2\end{pmatrix}\in\F[D]^{2\times 3}\text{ and }
    \widehat{G}=\begin{pmatrix}D+2&2+2D^2&D+2\end{pmatrix}\in\F[D]^{1\times 3}.
\]
It is easy to see that~$G$ and~$\widehat{G}$ are minimal and basic and satisfy $G\widehat{G}^{\sf T}=0$.
Thus, the codes $\cC:=\im G$ and $\widehat{\cC}:=\im\widehat{G}$ are mutual duals.
The CCF's of the given encoders~$G$ and~$\widehat{G}$ are
\[
  (A,B,C,E)=\left(\begin{pmatrix}0&1\\0&0\end{pmatrix},\,\begin{pmatrix}1&0\\0&0\end{pmatrix},\,
                  \begin{pmatrix}0&1&0\\1&0&0\end{pmatrix},\,\begin{pmatrix}1&2&0\\1&0&2\end{pmatrix}\right)
\]
and
\[
  (\widehat{A},\widehat{B},\widehat{C},\widehat{E})
    =\left(\begin{pmatrix}0&1\\0&0\end{pmatrix},\,\begin{pmatrix}1&0\end{pmatrix},\,
                  \begin{pmatrix}1&0&1\\0&2&0\end{pmatrix},\,\begin{pmatrix}2&2&2\end{pmatrix}\right),
\]
respectively. Using the lexicographic ordering of the states in $\F^2$
\begin{equation}\label{e-ordering}
  (0,0),\,(0,1),\,(0,2),\,(1,0),\,(1,1),\,(1,2),\,(2,0),\,(2,1),\,(2,2),
\end{equation}
the associated WAM of~$G$ is given by
{\small
\[
  \Lambda=\begin{pmatrix}
            1\!+2W^2\!\!\!\!&0&0&\!\!\!\!\!\!2W^2\!+W^3\!\!\!\!&0&0&\!\!\!\!\!\!2W^2\!+W^3\!\!\!\!&0&0\\
            2W\!+W^2\!\!\!\!&0&0&\!\!\!\!\!\!2W^2\!+W^3\!\!\!\!&0&0&\!\!\!\!\!\!W\!+2W^3\!\!\!\!&0&0\\
            2W\!+W^2\!\!\!\!&0&0&\!\!\!\!\!\!W\!+2W^3\!\!\!\!&0&0&\!\!\!\!\!\!2W^2\!+W^3\!\!\!\!&0&0\\
            0&\!\!\!\!W\!+2W^3\!\!\!\!&0&0&\!\!\!\!2W\!+W^2\!\!\!\!&0&0&\!\!2W^2\!+W^3\!\!\!\!&0\\
            0&\!\!\!\!2W^2\!+W^3\!\!\!\!&0&0&\!\!\!\!2W\!+W^2\!\!\!\!&0&0&\!\!W\!+2W^3\!\!\!\!&0\\
            0&\!\!\!\!2W^2\!+W^3\!\!\!\!&0&0&\!\!\!\!1\!+2W^2\!\!\!\!&0&0&\!\!2W^2\!+W^3\!\!\!\!&0\\
            0&0&\!\!\!\!W\!+2W^3\!\!&0&0&\!\!\!\!2W^2\!+W^3\!\!\!\!&0&0&\!\!\!\!2W\!+W^2\\
            0&0&\!\!\!\!2W^2\!+W^3\!\!\!\!&0&0&\!\!\!\!2W^2\!+W^3\!\!\!\!&0&0&\!\!\!\!1\!+2W^2\\
            0&0&\!\!\!\!2W^2\!+W^3\!\!\!\!&0&0&\!\!\!\!W\!+2W^3\!\!\!\!&0&0&\!\!\!\!2W\!+W^2\end{pmatrix}.
\]
}
\hspace*{-.37em}For instance, the entry at position $(6,2)$ is obtained as follows.
Since the $6$th state is $X=(1,2)$ and the $2$nd state is $Y=(0,1)$ we have to consider the outputs
$v=(1,\,2)C+(u_1,\,u_2)E=(2+u_1+u_2,\,1+2u_1,\,2u_2)$, where $(u_1,u_2)\in\F^2$ is such that
$(0,\,1)=(1,\,2)A+(u_1,\,u_2)B$.
Since $(1,\,2)A+(u_1,\,u_2)B=(u_1,\,1)$ this is the case if and only if $u_1=0$ and we see that the entry
at position~$(6,2)$ is given by
$\we\{(2+u_2,\,1,\,2u_2)\mid u_2=0,1,2\}=\we\{(2,1,0),\,(0,1,2),\,(1,1,1)\}=2W^2+W^3$.
Likewise the WAM associated with the state space realization
$(\widehat{A},\widehat{B},\widehat{C},\widehat{E})$ of~$\widehat\cC$ can be computed as
\begin{equation}\label{e-Lambdahat}
  \widehat{\Lambda}=\begin{pmatrix}
            1&0&0&W^3&0&0&W^3&0&0\\
            W&0&0&W^3&0&0&W^2&0&0\\
            W&0&0&W^2&0&0&W^3&0&0\\
            0&W^2&0&0&W&0&0&W^3&0\\
            0&W^3&0&0&W&0&0&W^2&0\\
            0&W^3&0&0&1&0&0&W^3&0\\
            0&0&W^2&0&0&W^3&0&0&W\\
            0&0&W^3&0&0&W^3&0&0&1\\
            0&0&W^3&0&0&W^2&0&0&W\end{pmatrix}.
\end{equation}
As one can see, the WAM contains a considerable amount of redundancy in its entries.
For further details we refer to \cite[Sec.~III]{GS08}.
\end{exa}

It is clear from Definition~\ref{D-Lambda} that the WAM depends on the chosen
encoder~$G$.
In order to describe this dependence we associate with any $P\in \mbox{\rm{GL}}_{\delta}(\F)$ the
permutation matrix
\begin{equation}\label{e-PP}
  \cP(P)\in \mbox{\rm{GL}}_{q^{\delta}}(\C),
  \text{ where $\cP(P)_{X,Y}=1$ if $Y=XP$ and $\cP(P)_{X,Y}=0$ else.}
\end{equation}
Furthermore, let $\Pi:=\{\cP(P)\mid P\in \mbox{\rm{GL}}_{\delta}(\F)\}$ denote the group of all
such permutation matrices.
By definition, the matrix $\cP(P)$ corresponds to the permutation on the set $\F^{\delta}$
induced by the isomorphism~$P$.
Obviously, for any matrix $\Lambda\in\C[W]^{q^{\delta}\times q^{\delta}}$ and any
$\cP:=\cP(P)\in\Pi$ we have
\begin{equation}\label{e-Pitrafo}
     \big(\cP\Lambda\cP^{-1}\big)_{X,Y}=\Lambda_{XP,YP}\text{ for all }(X,Y)\in\F^{\delta}\times\F^{\delta}.
\end{equation}
In \cite[Thm.~4.1]{GL05p} it has been shown that for a given code~$\cC$
\begin{equation}\label{e-Lambdaunique}
 \left.\begin{array}{ll} G_1,\,G_2\in\F[D]^{k\times n}\\
 \text{are minimal encoders of }\cC\end{array}\right\}\Longrightarrow
 \Lambda(G_1)=\cP\Lambda(G_2)\cP^{-1}\text{ for some }\cP\in\Pi.
\end{equation}
As a consequence, for a given code~$\cC$ with minimal encoder~$G$ and WAM $\Lambda=\Lambda(G)$ the
equivalence class
\begin{equation}\label{e-genLambda}
 [\Lambda]:=\{\cP\Lambda\cP^{-1}\mid \cP\in\Pi\}
\end{equation}
forms an invariant of the code.
It is called the {\em generalized WAM\/} of~$\cC$.

For the rest of this paper we will fix the following data.

\begin{ass}\label{A-data1}
Let $\F=\F_q$ be a field with~$q=p^s$ elements.
Let $\cC\subseteq\F[D]^n$ be an $(n,k,\delta)$ code with~$r$ nonzero Forney indices and define
$\cF:=\F^{\delta}\times\F^{\delta}$.
Furthermore, let $G\in\F[D]^{k\times n}$ be a minimal encoder of~$\cC$ with the first~$r$ rows
corresponding to the nonzero Forney indices.
Let $(A,B,C,E)$ be the corresponding CCF and $\Lambda$ be the associated WAM.
\\
Likewise, let the dual code $\widehat\cC$ have~$\widehat{r}$ nonzero Forney indices and let
$\widehat G\in\F[D]^{(n-k)\times n}$ be a minimal encoder with the first~$\widehat{r}$ rows
corresponding to the nonzero Forney indices.
Let $(\widehat A,\widehat B,\widehat C, \widehat E)$ be the corresponding CCF and
denote the associated WAM by~$\widehat{\Lambda}$.
\end{ass}

Recall from\eqnref{e-Cperp} that~$\cC$ and~$\widehat\cC$ both have degree~$\delta$ and thus the
WAM's $\Lambda$ and $\widehat\Lambda$ are both in $\C[W]^{q^{\delta}\times q^{\delta}}$.

In order to formulate our main result, we need to introduce a certain transformation matrix.
They are well known from the MacWilliams Identity for the complete weight enumerator for block codes.
Choose a primitive $p$-th root of unity $\zeta\in\C^*$ and consider the trace form
$\tau:\F_q\longrightarrow\F_p,\ a\longmapsto\sum_{i=0}^{s-1}a^{p^i}$.
Then we define the MacWilliams matrix to be
\begin{equation}\label{e-cH}
   \cH:=q^{-\frac{\delta}{2}}\Big(\zeta^{\tau(XY^{\sf T})}\Big)_{X,Y\in\F^{\delta}}
   \in\C^{q^{\delta}\times q^\delta}.
\end{equation}

Now we are ready to present our main result.

\begin{theorem}\label{T-MacWID}
Let~$\cC$ and~$\widehat{\cC}$ and the associated data be as in General Assumption~\ref{A-data1}.
Then there exists some $P\in \mbox{\rm{GL}}_{\delta}(\F)$ such that
\begin{equation}\label{e-MacW}
   \widehat\Lambda_{X,Y}=q^{-k}\HH\big((\cH\Lambda^{\sf T}\cH^{-1})_{XP,YP}\big)
   \text{ for all }(X,Y)\in\cF,
\end{equation}
where~$\HH$ is as in\eqnref{e-h}.
As a consequence, the generalized WAM's satisfy
\begin{equation}\label{e-genMacWId}
  [\widehat{\Lambda}]=q^{-k}\HH(\cH[\Lambda]^{\sf T}\cH^{-1}).
\end{equation}
In other words, the matrix $q^{-k}\HH\big(\cH\Lambda^{\sf T}\cH^{-1}\big)$ is a representative of the
generalized WAM of~$\widehat\cC$.
\end{theorem}

Recall that, due to\eqnref{e-Lambdaunique}, the WAM's for two different minimal
encoders of~$\widehat{\cC}$ differ by conjugation with a suitable matrix $\cP(P)\in\Pi$.
This explains the presence of the matrix $P\in \mbox{\rm{GL}}_{\delta}(\F)$ in\eqnref{e-MacW}.
Of course,~$P$ depends on the chosen encoders~$G$ and~$\widehat{G}$.
In terms of the generalized WAM's, however, no specific representation of the code and no
transformation matrix are needed anymore.
Note also that in the case where $\delta=0$, the identity\eqnref{e-genMacWId} immediately
leads to the MacWilliams identity for block codes as given in\eqnref{e-MacWBC}.

The proof of Theorem~\ref{T-MacWID} is rather technical and will be presented in the next section.
The resulting version including an explicit transformation matrix~$P$ will be summarized in
Theorem~\ref{T-MacWID}$\,'$ at the end of the next section.

\section{Proof of Theorem~\ref{T-MacWID}}\label{S-Proof}
\setcounter{equation}{0}
Let the data be as in General Assumption~\ref{A-data1}.
The following simple properties of the matrices in a CCF will come handy
throughout this section.

\begin{remark}\label{R-abceprop}
The matrices $(A,B,C,E)$ as in Definition~\ref{D-CCF} have the following properties.
\begin{romanlist}
\item $AB^{\sf T}=0,\ BB^{\sf T}B=B,\ AA^{\sf T}A=A$,
\item $\im B^{\sf T}=\im(I_r,\,0)\subseteq\F^k$ and $\ker B=\im (0,\,I_{k-r})\subseteq\F^k$,
\item $\im A\cap \im B=\{0\}$,
\item $\cCconst=(\ker B)E:=\{uE\mid u\in\ker B\}\ \text{ and }\ \im E=\im B^{\sf T}E\oplus\cCconst$,
\item $\ker A\cap \ker C=\{0\}$,
\item $(\ker A)C\cap\cCconst=\{0\}$,
\end{romanlist}
The first~4~properties are easily verified, see also \cite[Rem.~II.4, Rem.~II.6]{GS08}.
The last two properties are due to the fact that the encoder matrix $G=B(D^{-1}I-A)^{-1}C+E$ is minimal.
Indeed, notice that $\ker A=\text{span}_{\F}\{e_{j_l}\mid l=1,\ldots,r\}$, where $j_l=\sum_{i=1}^l\delta_i$, and where
$e_1,\ldots,e_{\delta}$ denote the standard basis vectors in~$\F^{\delta}$.
Using~$G$ as in Definition~\ref{D-CCF} we see that $e_{j_l}C=g_{l,\delta_l}$, the highest coefficient
vector of the $l$th row of~$G$.
Recalling that for a minimal matrix~$G$ the highest coefficient vectors
$g_{1,\delta_1},\ldots,g_{k,\delta_k}$ are linearly independent and
noticing that $\cCconst=\text{span}_{\F}\{g_{r+1,\delta_{r+1}},\ldots,g_{k,\delta_k}\}$, one easily derives
properties~(v) and~(vi).
\end{remark}

In the sequel the spaces
\begin{equation}\label{e-DeltaOmega}
   \Delta:=\im\begin{pmatrix}I&A\\0&B\end{pmatrix} \text{ and }\
   \Omega:=\big\{(X,Y)\in\Delta\,\big|\, XC+YB^{\sf T}E\in\cCconst\big\}
\end{equation}
will play a crucial role.
Notice that
\begin{equation}\label{e-DeltaLambda}
   \Delta=\{(X,Y)\in\cF\mid Y=XA+uB\text{ for some }u\in\F^k\}=\{(X,Y)\in\cF\mid\Lambda_{X,Y}\neq 0\},
\end{equation}
that is,~$\Delta$ is the space of all ordered pairs of states $(X,Y)$ admitting a direct transition
$Y=XA+uB$ for some suitable input~$u$.
The set~$\Omega$ describes those state pairs for which one of the transitions leads to zero output.
Indeed, we have

\begin{proposition}\label{P-Omega}
$\Omega=\{(X,Y)\in\Delta\mid \exists\; u\in\F^k:\ Y=XA+uB,\ 0=XC+uE\}$.
\end{proposition}
\begin{proof}
We will make use of Remark~\ref{R-abceprop}(i) and~(iv).
For ``$\subseteq$'' let $(X,Y)\in\Omega$.
Then $Y=XA+uB$ for some $u\in\F^k$ and we compute
$XC+YB^{\sf T}E=XC+XAB^{\sf T}E+uBB^{\sf T}E=XC+uBB^{\sf T}E$.
Since, by assumption, this vector is in $\cCconst$, we obtain
$XC+uBB^{\sf T}E=u'E$ for some $u'\in\ker B$.
Now $XC+(uBB^{\sf T}-u')E=0$ and therefore $(X,Y)=(X,XA+uB)=(X,XA+(uBB^{\sf T}-u')B)$
is in the set on the right hand side.
\\
For ``$\supseteq$'' let $Y=XA+uB$ and $0=XC+uE$.
Using Remark~\ref{R-abceprop}(i) and~(iv) we compute
\begin{align*}
  XC+(XA+uB)B^{\sf T}E&=XC+uBB^{\sf T}E=XC+uE+ u(BB^{\sf T}-I)E\\
                      &=u(BB^{\sf T}-I)E\in(\ker B)E=\cCconst.
\end{align*}
As a consequence, $(X,Y)\in\Omega$.
\end{proof}

The following results will be crucial.
The first three statements are easily obtained from the form of the matrices $(A,B,C,E)$ and can be found in
\cite[Prop.~III.6, Lem.~III.7, Prop.~III.11]{GS08}.
The last result needs some more detailed considerations and has been proven in \cite[Lem.~III.9]{GS08},
where the space~$\Omega$ appears as $\ker\Phi$.

\begin{proposition}\label{P-Delta}\
\begin{alphalist}
\item $\dim\Delta=\delta+r$.
\item The orthogonal of~$\Delta$ in~$\cF$ is given by
      $\Delta^\perp=\{(XA^{\sf T},-XA^{\sf T}A)\mid X\in \F^\delta\}$.
\item $\Delta\oplus\Delta^-=\cF$, where $\Delta^-:=\{(0,Y)\mid Y\in \im A\}$.
\item $\dim\Omega=\delta-\widehat{r}$, where~$\widehat{r}$ is as in General Assumption~\ref{A-data1}.
\end{alphalist}
\end{proposition}

In the paper~\cite{GS08}, a weak version of the MacWilliams Identity~\ref{T-MacWID} has been established.
In order to present that result, we define
\begin{equation}\label{e-M0}
  M_0:=\begin{pmatrix}\widehat{C}C^{\sf T}&\widehat{C}E^{\sf T}B\\ \widehat{B}^{\sf T}\widehat{E}C^{\sf T}&0
       \end{pmatrix}
  \in\F^{2\delta\times2\delta}\ \text{ and } \cM_0:=\im M_0.
\end{equation}
Let us also consider the dual versions of the spaces in\eqnref{e-DeltaOmega} and
Proposition~\ref{P-Delta}(c);
that is, let $\widehat{\Delta},\; \widehat{\Omega},$ and $\widehat{\Delta}^-$ denote the respective spaces associated
with the dual code $\widehat{\cC}$.
In the sequel we will make frequent use of the dual results of Proposition~\ref{P-Delta}.
From \cite[Lem.~V.3]{GS08} it follows that
\begin{equation}\label{e-M0prop}
  \widehat{\Omega}\oplus\widehat{\Delta}^-=\ker M_0,\quad \cM_0\subseteq\Omega^{\perp},\;\text{ and }\;
  \cM_0\oplus\Delta^{\perp}=\Omega^{\perp}.
\end{equation}
Notice that, according to Proposition~\ref{P-Delta}(a) and~(d) and their dual versions,
$\dim\widehat{\Omega}=\delta-r=\dim\Delta^{\perp}$.
Now let us choose a direct complement~$\widehat{\Delta}^*$ of $\widehat{\Omega}$ in~$\widehat{\Delta}$ and
let~$\cG$ be a direct complement of~$\Omega^{\perp}$ in~$\cF$.
Then $\dim\cG=2\delta-(\delta+\widehat{r})=2\delta-\dim\widehat{\Delta}=\dim\widehat{\Delta}^-$.
Furthermore, let $f_0:\,\widehat{\Delta}^*\longrightarrow\cM_0$ be the isomorphism
$(X,Y)\longmapsto (X,Y)M_0$.
All this leads to the diagram
\begin{equation}\label{e-diagram}
\begin{array}{l}
    \hspace*{4.3cm}\widehat\Delta\\[.2ex]
    \hspace*{2.8cm}\overbrace{\hspace*{3.3cm}}\\[-.3ex]
    \xymatrix{
    \cF\ar[d]_{f}&=\!\!&\widehat\Delta^*\ar[d]_{f_0}\!\!&\oplus\!\!\!\!&
                               \widehat{\Omega}\ar[d]_{f_1}&\oplus&\widehat\Delta^-\ar[d]_{f_2}\\
    \cF\,&=&\!\!\cM_0             &\!\!\oplus\!\!\!\!\!\!&\Delta^{\perp}   &\!\!\oplus\!\!&\cG
     }\\[-1.8ex]
    \hspace*{2.8cm}\underbrace{\hspace*{3.3cm}}\\[1ex]
    \hspace*{4.3cm}\Omega^{\perp}
\end{array}
\end{equation}
where, due to the dimensions, there exist vector space isomorphisms
$f_1:\widehat{\Omega}\longrightarrow\Delta^{\perp}$ and $f_2:\widehat{\Delta}^-\longrightarrow\cG$ in the
last two columns and where $f=f_0\oplus f_1\oplus f_2$.

Now we can present a cornerstone in the proof of the MacWilliams Identity.
The following weak version of the identity has been proven in \cite[Thm.~V.5]{GS08}.
It shows that~$\widehat\Lambda$ and $q^{-k}\HH\big(\cH\Lambda^{\sf T}\cH^{-1})$ have the same
entries up to an automorphism~$f$ on the space~$\cF$ of state pairs.

\begin{theorem}
\label{T-MacWstart}
Consider the diagram\eqnref{e-diagram}.
Then
\[
   \widehat\Lambda_{f^{-1}(-Y,X)} =q^{-k}\HH\big((\cH\Lambda^{\sf T}\cH^{-1})_{X,Y}\big)
   \text{ for all }(X,Y)\in\cF.
\]
where $f$ is the automorphism on~$\cF$ defined as $f:=f_0\oplus f_1\oplus f_2$.
\end{theorem}
It is worth being stressed that in this theorem the spaces $\widehat{\Delta}^*$ and~$\cG$ are any arbitrary
direct complements of $\widehat{\Omega}$ in~$\widehat{\Delta}$ and of~$\Omega^{\perp}$ in~$\cF$, respectively.
Also, the isomorphisms~$f_1$ and~$f_2$ are not further specified.
In order to prove our main result, Theorem~\ref{T-MacWID}, we will use this remaining freedom
such that the resulting automorphism $f=f_0\oplus f_1\oplus f_2$ is of the form as desired in
Theorem~\ref{T-MacWID}.
More precisely, we need~$f$ to respect the decomposition
$\cF=\F^{\delta}\times\F^{\delta}$, that is,
\begin{equation}\label{e-mapf}
    f(X,Y)=(X,Y)\begin{pmatrix}0&P\\-P&0\end{pmatrix} \text{ for all }(X,Y)\in\cF
\end{equation}
for some state space isomorphism $P\in\mbox{\rm{GL}}_{\delta}(\F)$.
Indeed, with~$f$ being of this form we obtain $f^{-1}(-Y,X)=(XP^{-1},YP^{-1})$ and
the identity in Theorem~\ref{T-MacWstart} turns into
$\widehat{\Lambda}_{XP^{-1},YP^{-1}}=q^{-k}\HH\big((\cH\Lambda^{\sf T}\cH^{-1})_{X,Y}\big)$ for all
$(X,Y)\in\cF$.
This is exactly the statement of Theorem~\ref{T-MacWID}.
The rest of this section will be devoted to specifying the choice of the spaces $\widehat{\Delta}^*$ and~$\cG$ as
well as the isomorphisms~$f_1$ and~$f_2$ in Diagram\eqnref{e-diagram} in order to meet the
requirement\eqnref{e-mapf}.

Let us begin with the following technical facts.

\newpage
\begin{proposition}\label{P-ChatS0}\
\begin{alphalist}
\item Let $\Pi_1:\;\cF\longrightarrow\F^{\delta}$ be the projection onto the first component,
      thus $\Pi_1(X,Y)=X$ for all $(X,Y)\in\cF$. Then $\Pi_1|_{\textstyle\Omega}$ is injective.
\item $\rank C\widehat{E}^{\sf T}\widehat{B}=\widehat{r}$.
\item $\ker C\widehat{E}^{\sf T}\widehat{B}=\Pi_1(\Omega)=
       \{X\in\F^{\delta}\mid \exists\; u\in\F^k:\; (X,XA+uB)\in\Omega\}$.
\end{alphalist}
\end{proposition}
\begin{proof}
(a) Suppose $(0,uB)\in\Omega$ for some $u\in\F^k$.
Then $uBB^{\sf T}E\in\cCconst$.
But then Remark~\ref{R-abceprop}(i) and (iv) along with the full row rank of~$E$ yield
$uB=uBB^{\sf T}B=0$, which proves~(a).
\\
(b)
Again we will employ Remark~\ref{R-abceprop}(i) and~(iv).
Let $X\in\F^{\delta}$ such that $XC\widehat{E}^{\sf T}\widehat{B}=0$.
Using\eqnref{e-CC} we have, on the one hand, $XC\in\cCcoeff=(\cCperpconst)^{\perp}$,
where the last identity is due to Proposition~\ref{P-duality}.
On the other hand,
$XC\in\ker(\widehat{E}^{\sf T}\widehat{B})=(\im \widehat{B}^{\sf T}\widehat{E})^{\perp}$.
Making use of Remark~\ref{R-abceprop}(iv) and its dual version this yields
$XC\in(\cCperpconst)^{\perp}\cap(\im\widehat{B}^{\sf T}\widehat{E})^{\perp}
  =(\cCperpconst\oplus\im \widehat{B}^{\sf T}\widehat{E})^{\perp}=(\im\widehat{E})^{\perp}$.
But the latter space is identical to $\im E$, as one can see directly from the identity
$0=G(D)\widehat{G}(D)\T=\big(B(D^{-1}I-A)^{-1}C+E\big)
       \big(\widehat{B}(D^{-1}I-\widehat{A})^{-1}\widehat{C}+\widehat{E}\big)\T$
and the full row rank of the matrices~$E$ and~$\widehat{E}$.
Thus we conclude that $XC\in\im E=\cCconst\oplus\im B^{\sf T}E$.
Using that $B^{\sf T}BB^{\sf T}=B^{\sf T}$, we obtain the existence of some $u=\tilde{u}B\T\in\F^k$ such that
$XC+uBB^{\sf T}E\in\cCconst$.
Along with the identity $AB^{\sf T}=0$ this implies that $(X,XA+uB)\in\Omega$.
All this shows that $\ker C\widehat{E}^{\sf T}\widehat{B}\subseteq\Pi_1(\Omega)$ and, using~(a),
we arrive at $\dim\ker C\widehat{E}^{\sf T}\widehat{B}\leq\dim\Pi_1(\Omega)=\dim\Omega=\delta-\widehat{r}$.
Since $C\widehat{E}^{\sf T}\widehat{B}\in\F^{\delta\times\delta}$
this implies $\widehat{r}\leq\rank C\widehat{E}^{\sf T}\widehat{B}\leq \rank\widehat{E}^{\sf T}\widehat{B}$.
Recalling from Proposition~\ref{P-duality} that $\dim\cCperpconst=n-k-\widehat{r}$, the dual version of
Remark~\ref{R-abceprop}(iv) along with $\rank\widehat{E}=n-k$
then tells us that $\rank\widehat{E}^{\sf T}\widehat{B}=\widehat{r}$.
This finally proves $\rank C\widehat{E}^{\sf T}\widehat{B}=\widehat{r}$.
\\
(c) The inclusion ``$\subseteq$'' has been shown in the proof of~(b).
Thus equality of the two spaces follows from Proposition~\ref{P-Delta}(d) since
$\dim\ker C\widehat{E}^{\sf T}\widehat{B}=\delta-\widehat{r}=\dim\Omega=\dim\Pi_1(\Omega)$.
\end{proof}

Part~(a) and~(c) of the previous proposition give rise to a crucial map.

\begin{corollary}\label{C-Khat}
Let $K:=\ker C\widehat{E}^{\sf T}\widehat{B}\subseteq\F^{\delta}$. Then
\[
   \sigma:\;K \longrightarrow\F^{\delta},\quad X\longmapsto Y \text{ such that }(X,Y)\in\Omega
\]
is a well-defined, linear, and injective map.
Furthermore, $K$ does not contain a nonzero $\sigma$-invariant subset.
\end{corollary}
\begin{proof}
Well-definedness follows from Proposition~\ref{P-ChatS0}(a) and~(c), whereas linearity is obvious.
As for injectivity, let $X\in K$ such that $\sigma(X)=0$.
Then $(X,0)\in\Omega$, meaning that $XC\in\cCconst$.
On the other hand, $(X,0)\in\Omega\subseteq\Delta$ tells us that $0=XA+uB$ for some $u\in\F^k$.
Hence $XA=-uB\in\im A\cap\im B$ and Remark~\ref{R-abceprop}(iii) implies $XA=0$.
But then $XC\in(\ker A)C\cap\cCconst=\{0\}$, where the last identity is due to Remark~\ref{R-abceprop}(vi).
As a consequence, $X\in\ker A\cap\ker C$, and due to~(v) of the same remark we arrive at $X=0$.
This proves the injectivity of~$\sigma$.
\\
For the last statement assume that~$K'$ is a $\sigma$-invariant subset of~$K$.
That simply means that there exists some vector $X\in K$ such that $\sigma^i(X)\in K$ for all $i\geq0$.
Since $K\subseteq\F^{\delta}$ is a finite set, this yields that the orbit $\{\sigma^i(X)\mid i\in\N_0\}$
is finite and hence contains a cycle.
In other words, there exists some $X'\in K$ and some $j>0$ such that $\sigma^j(X')=X'$.
Without loss of generality we may assume $X'=X$.
By definition of the map~$\sigma$ we have $\big(\sigma^i(X),\sigma^{i+1}(X)\big)\in\Omega$ for all $i\geq0$.
Using Proposition~\ref{P-Omega} all this tells us that we have a cycle
\[
   X\edge{u_0}{0}\sigma(X)\edge{u_1}{0}\sigma^2(X)\edge{u_2}{0}\cdots
   -\!\!\!-\!\!\!\longrightarrow\hspace{-2.6em}
   \raisebox{1.6ex}{\mbox{$\scriptscriptstyle (\!\frac{\,u_{j-1}\,}{0}\!)$}}\hspace{.5em}
   \sigma^j(X)=X
\]
of weight zero in the state transition diagram associated with $(A,B,C,E)$.
Here the notation $X\edge{u}{v}Y$ stands for the equations $Y=XA+uB,\,v=XC+uE$.
It is well-known \cite[p.~308]{LC83} that the basicness of the encoder~$G$ implies that
such a cycle is a concatenation of the trivial cycle, that is, $X=0$ and $u_i=0$ for all $i=0,\ldots,j-1$.
Thus $K'=\{0\}$ and the proof is complete.
\end{proof}

Let us now introduce the matrices
\begin{equation}\label{e-S}
   S_0:=B^{\sf T}E\ \text{ and }\ S_i:=B^{\sf T}BA^{i-1}C\text{ for }i\geq 1.
\end{equation}
Likewise, we define the matrices $\widehat{S}_0:=\widehat{B}^{\sf T}\widehat{E}$ and
$\widehat{S}_i:=\widehat{B}^{\sf T}\widehat{B}\widehat{A}^{i-1}\widehat{C},\,i\geq1$, associated with the dual
 code.
Furthermore, we put
\begin{equation}\label{e-N}
\begin{split}
    N&:=\sum_{m\geq 2}\sum_{i=1}^{m-1}\sum_{j=0}^{i-1}
           (\widehat A^{\sf T})^{i-1}\widehat S_jS_{m-j}^{\sf T}A^{m-(i+1)},\\
  \widehat{N}&:=\sum_{m\geq 2}\sum_{i=1}^{m-1}\sum_{j=0}^{i-1}
           (A^{\sf T})^{i-1}S_j\widehat{S}_{m-j}^{\sf T}\widehat{A}^{m-(i+1)}.
\end{split}
\end{equation}
Using that $A^i=0=\widehat{A}^i$ for $i\geq\delta$ it is easy to see that these sums are indeed finite
since each summand vanishes for $m\geq 2\delta$.
Using two index changes one easily shows that
\begin{equation}\label{e-Nhattrans}
  \widehat{N}^{\sf T}=\sum_{m\geq 2}\sum_{i=1}^{m-1}\sum_{j=i+1}^{m}
         (\widehat{A}^{\sf T})^{i-1} \widehat{S}_j S_{m-j}^{\sf T} A^{m-(i+1)}.
\end{equation}
In the appendix we prove the following technical, but straightforward properties.

\begin{proposition}\label{P-abce}\
\begin{alphalist}
\item $N+\widehat{N}^{\sf T}=-\widehat{C}C^{\sf T}$.
\item $\widehat{C}S_0^{\sf T}+\widehat{S}_0C^{\sf T}=NA+\widehat{A}^{\sf T}\widehat{N}^{\sf T}$.
\item $NAA^{\sf T}=N$.
\end{alphalist}
\end{proposition}

Now we are in a position to work on the remaining freedom in Diagram\eqnref{e-diagram}.
Define the matrices
\begin{equation}\label{e-Mmatrices}
  M_1=\begin{pmatrix}N& -NA\\0&0\end{pmatrix},\quad
  M_2=\begin{pmatrix}\widehat{N}^{\sf T}&0\\-\widehat{A}^{\sf T}\widehat{N}^{\sf T}&0 \end{pmatrix}.
\end{equation}
Recalling that $S_0=B^{\sf T}E$, Proposition~\ref{P-abce} along with the matrix~$M_0$ defined in\eqnref{e-M0}
yields
\begin{equation}\label{e-P}
  M:=M_0+M_1+M_2=\begin{pmatrix}0&P\\-P&0\end{pmatrix},\ \text{ where }
  P:=\widehat{C}S_0^{\sf T}-NA.
\end{equation}

In the rest of this section we will show that, firstly, the map~$f$ induced by $M_0+M_1+M_2$ is an automorphism,
that is, the matrix~$P\in\F^{\delta\times\delta}$ is regular, and, secondly, that~$f$ respects the
decomposition of~$\cF$ on the right hand side of Diagram\eqnref{e-diagram}.
As a consequence,~$f$ defines an automorphism as in Theorem~\ref{T-MacWstart} that, at the same time, is of the
form as in\eqnref{e-mapf}.
All this will establish Theorem~\ref{T-MacWID}.

In order to carry out these computations notice that $M_2=\widehat{M}_1^{\,\sf T}$, i.~e.,~$M_2$ is the
dual version of $M_1^{\sf T}$.
From Remark~\ref{R-abceprop}(i) and Proposition~\ref{P-abce}(c) we obtain
\[
   \begin{pmatrix}N&-NA\\0&0\end{pmatrix}\begin{pmatrix}I&0\\A^{\sf T}&B^{\sf T}\end{pmatrix}=0.
\]
Consequently,
\begin{equation}\label{e-M1M2}
    \im M_1\subseteq\Delta^{\perp}\ \text{ and }\ \widehat{\Delta}\subseteq\ker M_2,
\end{equation}
where the second containment follows from the first one via duality.
The following result establishes the regularity of~$P$.

\begin{theorem}\label{T-Pmatrix}
The matrix~$P=\widehat{C}S_0^{\sf T}-NA$ is in $\mbox{\rm{GL}}_{\delta}(\F)$.
\end{theorem}
\begin{proof}
We need to resort to the dual version of Corollary~\ref{C-Khat}.
Thus, consider $\widehat{K}=\ker\widehat{C}E^{\sf T}B$ with the corresponding map $\widehat{\sigma}$.
Firstly, one observes that $\ker P \subseteq\widehat{K}$. Indeed, for $X\in\ker P$ we have
$XNA=X\widehat{C}S_0^{\sf T}=X\widehat{C}E^{\sf T}B\in\im A\cap \im B$,
and from Remark~\ref{R-abceprop}(iii) we conclude
\begin{equation}\label{e-XNA}
   XNA=X\widehat{C}E^{\sf T}B=0.
\end{equation}
This proves $\ker P\subseteq\widehat{K}$.
In order to show the regularity of~$P$, let $X\in\ker P$. Then $X\in\widehat{K}$ and thus
$(X,\widehat{\sigma}(X))\in\widehat{\Omega}$.
Recalling that $\widehat{\Omega}\subseteq\widehat{\Delta}$ we obtain from\eqnref{e-M1M2} and\eqnref{e-M0prop}
\begin{equation}\label{e-kerM0M2}
   (X,\widehat{\sigma}(X))\in\ker M_2\cap\ker M_0.
\end{equation}
Moreover, $(X,\widehat{\sigma}(X))M_1=(XN,-XNA)$.
But $XNA=0$ by\eqnref{e-XNA} and thus Proposition~\ref{P-abce}(c) yields $XN=XNAA\T=0$.
Hence $(X,\widehat{\sigma}(X))\in\ker M_1$, which along with\eqnref{e-kerM0M2} implies
$(X,\widehat{\sigma}(X))\in\ker M$.
Consequently, $\widehat{\sigma}(X)\in\ker P$.
All this shows that $\ker P$ is a $\widehat{\sigma}$-invariant subspace of $\widehat{K}$, and by the dual
version of Corollary~\ref{C-Khat} we may conclude that $\ker P=\{0\}$.
This yields the desired result.
\end{proof}

This theorem shows that the map~$f$ induced by $M=M_0+M_1+M_2$ is an automorphism on~$\cF$ of the form
as in\eqnref{e-mapf}.
In order to complete the proof of Theorem~\ref{T-MacWID} it only remains to show that~$f$ is as in
Theorem~\ref{T-MacWstart}, that is, that it respects the direct decomposition as in Diagram\eqnref{e-diagram}.
This is accomplished and summarized in the next result.

\begin{proposition}\label{P-Mmatrices}
Put $\widehat{\Delta}^*:=\ker M_1\cap \widehat{\Delta}$ and $\cG:=\im M_2$.
Then
\begin{alphalist}
\item $\ker M_0=\widehat{\Omega}\oplus\widehat{\Delta}^-$.
\item $\ker M_1=\widehat{\Delta}^*\oplus\widehat{\Delta}^-$.
\item $\ker M_2=\widehat{\Delta}^*\oplus\widehat{\Omega}=\widehat{\Delta}$.
\item $\im M_1=\Delta^{\perp}$ and $\cF=\Omega^{\perp}\oplus\cG$.
\end{alphalist}
\end{proposition}
\begin{proof}
(a) has already been given in\eqnref{e-M0prop}.
\\
(b) It is clear from the definition of $\widehat{\Delta}^*$ and the dual version of Proposition~\ref{P-Delta}(c)
that the sum $\widehat{\Delta}^*\oplus\widehat{\Delta}^-$ is indeed direct and contained in $\ker M_1$.
In order to show equality let us first compute the rank of~$M_1$.
To this end we show that
\begin{equation}\label{e-OmegaM1}
   \widehat{\Omega}\cap \ker M_1=\{0\}.
\end{equation}
Due to\eqnref{e-M1M2} and part~(a) we have that $\widehat{\Omega}\subseteq\ker M_0\cap\ker M_2$.
Then $(X,Y)M_1=(X,Y)M$ for $(X,Y)\in\widehat{\Omega}$.
Now the regularity of the matrix~$M$, see Theorem~\ref{T-Pmatrix}, implies\eqnref{e-OmegaM1}.
Using the dual version of Proposition~\ref{P-Delta}(d) as well as\eqnref{e-M1M2}, we
conclude $\delta-r=\dim\widehat{\Omega}\leq\rank M_1\leq\dim\Delta^{\perp}$.
Since $\dim\Delta^{\perp}=\delta-r$ due to Proposition~\ref{P-Delta}(a),
this proves
\begin{equation}\label{e-rkM1}
   \rank M_1=\delta-r
\end{equation}
and
\begin{equation}\label{e-imM1}
   \im M_1=\Delta^{\perp}.
\end{equation}
Next we show that
\begin{equation}\label{e-DeltahatOmegahat}
   \widehat{\Delta}=\widehat{\Delta}^*\oplus\widehat{\Omega}.
\end{equation}
The directness of the sum on the right hand side as well as the inclusion ``$\supseteq$'' are obvious,
see also\eqnref{e-OmegaM1}.
Furthermore, notice that $\ker M_1+\widehat{\Delta}=\cF$ as $\widehat{\Delta}^-\subseteq\ker M_1$ and
$\widehat{\Delta}^-\oplus\widehat{\Delta}=\cF$.
Since $\ker M_1\cap\widehat{\Delta}=\widehat{\Delta}^*$, we obtain
with the aid of Proposition~\ref{P-Delta} that
$\dim\widehat{\Delta}^*=\dim(\ker M_1)+\dim\widehat{\Delta}-\dim\cF=r+\widehat{r}
 =\dim\widehat{\Delta}-\dim\widehat{\Omega}$.
All this proves\eqnref{e-DeltahatOmegahat}.
Along with the dual version of Proposition~\ref{P-Delta}(c) we arrive at
\begin{equation}\label{e-Fdecomp}
  \cF=\widehat{\Delta}^*\oplus\widehat{\Omega}\oplus\widehat{\Delta}^-,
\end{equation}
which is exactly the decomposition of~$\cF$ as in the upper row of Diagram\eqnref{e-diagram}.
Now we compute
$\dim(\ker M_1)=\delta+r=2\delta-\dim\widehat{\Omega}=\dim(\widehat{\Delta}^*\oplus\widehat{\Delta}^-)$,
which along with $\widehat{\Delta}^*\oplus\widehat{\Delta}^-\subseteq\ker M_1$
completes the proof of~(b).
\\
(c)
Due to\eqnref{e-DeltahatOmegahat} it only remains to show that $\widehat{\Delta}=\ker M_2$.
The inclusion ``$\subseteq$'' has been obtained in\eqnref{e-M1M2}.
In order to establish identity recall that $M_2=\widehat{M}_1^{\,\sf T}$ and therefore dualizing\eqnref{e-rkM1}
yields $\rank M_2=\delta-\widehat{r}$.
But then $\dim(\ker M_2)=\delta+\widehat{r}=\dim\widehat{\Delta}$.
Hence $\ker M_2=\widehat{\Delta}$, which concludes the proof of~(c).
\\
(d) The first part has already been proven in\eqnref{e-imM1} above.
Furthermore, from~(c) we know that $\dim\cG=\dim(\im M_2)=\delta-\widehat{r}$.
Moreover, $\dim\Omega^{\perp}=\delta+\widehat{r}$.
Hence the proof of~(d) is complete if we can show that $\cG\cap\Omega^{\perp}=\{0\}$.
To this end assume $(X,Y)M_2\in\Omega^{\perp}$ for some $(X,Y)\in\cF$.
By~(c) and\eqnref{e-Fdecomp} we may assume $(X,Y)\in\widehat{\Delta}^-$ and therefore
$(X,Y)M_2=(X,Y)M$ due to~(a) and~(b).
Furthermore, by\eqnref{e-M0prop} we have $\Omega^{\perp}=\im M_0\oplus\Delta^{\perp}=\im M_0\oplus\im M_1$.
As a consequence, the above yields
$(X,Y)M_2=(X_0,Y_0)M_0+(X_1,Y_1)M_1$ for some $(X_i,Y_i)\in\cF,\,i=1,2$.
Using\eqnref{e-Fdecomp} and~(a) and~(b) we may assume $(X_0,Y_0)\in\widehat{\Delta}^*$ and
$(X_1,Y_1)\in\widehat{\Omega}$.
Using once more~(a) --~(c) we conclude
$(X,Y)M=(X,Y)M_2=(X_0,Y_0)M_0+(X_1,Y_1)M_1=(X_0,Y_0)M+(X_1,Y_1)M$ and regularity of the matrix~$M$ implies
$(X,Y)=(X_0,Y_0)+(X_1,Y_1)\in\widehat{\Delta}^-\cap(\widehat{\Delta}^*\oplus\widehat{\Omega})
=\widehat{\Delta}^-\cap\widehat{\Delta}$.
Thanks to Proposition~\ref{P-Delta}(c) this intersection is trivial and we may finally conclude that
$\cG\cap\Omega^{\perp}=\{0\}$.
This completes the proof.
\end{proof}

The proposition shows that the space~$\cF$ decomposes exactly as in Diagram\eqnref{e-diagram}
and that the matrices $M_i,\,i=0,1,2$, induce isomorphisms $f_i,i=0,1,2$.
As outlined in the paragraph right after\eqnref{e-mapf},
Theorem~\ref{T-MacWstart} along with\eqnref{e-mapf},\eqnref{e-P} and
Theorem~\ref{T-Pmatrix} conclude the proof of Theorem~\ref{T-MacWID}.
We summarize the result as follows.

\noindent{\bf Theorem~\ref{T-MacWID}$\,'$}
{\sl
Let~$\cC$ and~$\widehat{\cC}$ and the associated data be as in General Assumption~\ref{A-data1}.
Put $P:=\widehat{C}E^{\sf T}B-NA$, where $N$ is as in\eqnref{e-N}.
Then $P\in\mbox{\rm{GL}}_{\delta}(\F)$ and the WAM's of~$\cC$ and~$\widehat{\cC}$ satisfy the MacWilliams Identity
\[
   \widehat\Lambda_{X,Y}=q^{-k}\HH\big((\cH\Lambda^{\sf T}\cH^{-1})_{XP,YP}\big)
   \text{ for all }(X,Y)\in\cF.
\]
Consequently, the generalized WAM's $[\Lambda]$ and $[\widehat{\Lambda}]$ of~$\cC$
and~$\widehat{\cC}$ satisfy $[\widehat{\Lambda}]=q^{-k}\HH(\cH[\Lambda]^{\sf T}\cH^{-1})$.
}

We close this section with illustrating the MacWilliams Identity for the code in Example~\ref{E-1}.

\begin{exa}\label{E-2}
Let the codes $\cC=\im G$ and $\widehat{\cC}=\im\widehat{G}$ be as in Example~\ref{E-1}.
In order to carry out the transformation $q^{-k}\HH\big(\cH\Lambda^{\sf T}\cH^{-1}\big)$, we need the
MacWilliams matrix~$\cH$.
With the same ordering of the states as in\eqnref{e-ordering} one obtains
\[
  \cH=\frac{1}{3}\begin{pmatrix}
   1& 1& 1& 1& 1& 1& 1& 1& 1\\
   1& \zeta& \zeta^2& 1& \zeta& \zeta^2& 1& \zeta& \zeta^2\\
   1& \zeta^2& \zeta& 1& \zeta^2& \zeta& 1& \zeta^2& \zeta\\
   1& 1& 1& \zeta& \zeta& \zeta& \zeta^2& \zeta^2& \zeta^2\\
   1& \zeta& \zeta^2& \zeta& \zeta^2& 1& \zeta^2& 1& \zeta\\
   1& \zeta^2& \zeta& \zeta& 1& \zeta^2& \zeta^2& \zeta& 1\\
   1& 1& 1& \zeta^2& \zeta^2& \zeta^2& \zeta& \zeta& \zeta\\
   1& \zeta& \zeta^2& \zeta^2& 1& \zeta& \zeta& \zeta^2& 1\\
   1& \zeta^2& \zeta& \zeta^2& \zeta& 1& \zeta& 1& \zeta^2\end{pmatrix},
   \text{ where }\zeta=e^{\frac{2\pi i}{3}}.
\]
Now we may start computing the right hand side of the MacWilliams identity in Theorem~\ref{T-MacWID}'.
Using the matrix~$\Lambda$ from Example~\ref{E-1} we obtain
$\cH\Lambda^{\sf T}\cH^{-1}=\Gamma$, where
\[
  \Gamma=
  \begin{pmatrix}f_1& 0& 0& 0& f_4& 0& 0& 0& f_4\\
 0& 0& f_4& f_1& 0& 0& 0 & f_4& 0\\
 0& f_4& 0& 0& 0& f_4&f_1& 0& 0\\
 0& 0& f_3& f_2& 0& 0& 0& f_4& 0\\
 0& f_4& 0& 0& 0& f_3& f_2& 0& 0\\
 f_2& 0& 0& 0& f_4& 0& 0& 0& f_3\\
 0& f_3& 0& 0& 0& f_4 & f_2& 0& 0\\
 f_2& 0& 0& 0& f_3& 0& 0& 0& f_4\\
 0& 0& f_4& f_2& 0& 0& 0 & f_3& 0\end{pmatrix} \text{ with }
 \left\{\!\!\begin{array}{l}
   f_1\!=\frac{8}{3}W^3+4W^2+2W+\frac{1}{3},\\
   f_2\!=-\frac{4}{3}W^3+W+\frac{1}{3},\\
   f_3\!=\frac{2}{3}W^3-W^2+\frac{1}{3},\\
   f_4\!=-\frac{1}{3}W^3+W^2-W+\frac{1}{3}.\end{array}\right.
\]
Indeed, using that $\zeta^2+\zeta+1=0$, it can be checked straightforwardly that $\Gamma\cH=\cH\Lambda^{\sf T}$.
Next one easily computes the MacWilliams transforms $3^{-2}\HH(f_i)=W^{i-1}$ for $i=1,\ldots,4$, where~$\HH$ is as
in\eqnref{e-h}, and thus one obtains
\[
  \Phi:=3^{-2}\HH(\Gamma)=
  \begin{pmatrix}1& 0& 0& 0& W^3& 0& 0& 0& W^3\\
         0& 0& W^3& 1& 0& 0& 0& W^3& 0\\
         0& W^3& 0& 0& 0& W^3& 1& 0& 0\\
         0& 0& W^2& W& 0& 0& 0& W^3& 0\\
         0& W^3& 0& 0& 0& W^2& W& 0& 0\\
         W& 0& 0& 0& W^3& 0& 0& 0& W^2\\
         0& W^2& 0& 0& 0& W^3& W& 0& 0\\
         W& 0& 0& 0& W^2& 0& 0& 0& W^3\\
         0& 0& W^3& W& 0& 0& 0& W^2& 0\end{pmatrix}.
\]
Finally, we need to apply the state space isomorphism induced by the matrix~$P=\widehat{C}E^{\sf T}B-NA$.
Since $\delta=2$ the matrix~$N$ from\eqnref{e-N} is given by
\[
  N=\widehat{S}_0S_2^{\sf T}+\widehat{S}_0S_3^{\sf T}A+\widehat{A}^{\sf T}\widehat{S}_0S_3^{\sf T}
    +\widehat{A}^{\sf T}\widehat{S}_1S_2^{\sf T}
   =\widehat{S}_0S_2^{\sf T}+\widehat{A}^{\sf T}\widehat{S}_1S_2^{\sf T}
   =\begin{pmatrix}2&0\\1&0\end{pmatrix}.
\]
This yields $P=\widehat{C}E^{\sf T}B-NA=\Smallfourmat{1}{1}{1}{2}$.
Now one can check straightforwardly that $\Phi_{XP,YP}=\widehat{\Lambda}_{X,Y}$ for all
$(X,Y)\in\cF$, where~$\widehat{\Lambda}$ is the WAM of the dual code given in\eqnref{e-Lambdahat}.
This is exactly the identity in Theorem~\ref{T-MacWID}$\,'$.
\end{exa}

\section{Sequence Space Duality}\label{S-sequence}
\setcounter{equation}{0}
In this section we will briefly discuss a different notion of duality for CC's and
translate the MacWilliams Identity to this type of duality.

Thus, throughout this section let us call the dual~$\widehat{\cC}$ of a code
$\cC\subseteq\F[D]^n$ as defined in\eqnref{e-Cperpdef} the {\em module-theoretic dual}.
The literature on convolutional coding theory has also seen a notion of duality based on the $\F$-bilinear form
\[
   \F[D]^n\times\F[D]^n\longrightarrow \F,\quad
   \Big(\sum_{t\geq 0}v_tD^t,\,\sum_{t\geq 0}w_tD^t\Big)\longmapsto
   \langle\langle\sum_{t\geq 0}v_tD^t,\,\sum_{t\geq 0}w_tD^t\rangle\rangle:=\sum_{t\geq0}v_tw_t^{\sf T}.
\]
Notice that this sum is indeed finite since the vectors are polynomial.
The dual based on this bilinear form is usually, and most conveniently, defined in the setting of Laurent series,
see, e.~g.,~\cite{Fo01},~\cite{FT04},~\cite{JSW00},~\cite{Mi95}.
But we can just as well stay within our polynomial setting.
Then it amounts to defining the dual of the code~$\cC$ as
\begin{equation}\label{e-sdual}
   \widetilde{\cC}:=\{w\in\F[D]^n\mid \ddual{v,\,D^lw}=0\text{ for all }v\in\cC\text{ and }l\in\N_0\}.
\end{equation}
We call~$\widetilde{\cC}$ the {\em sequence space dual\/} of~$\cC$.
It is easy to see that~$\widetilde{\cC}$ is a submodule of $\F[D]^n$.
Furthermore, there is a simple relation between the sequence space dual~$\widetilde{\cC}$ and the
module-theoretic dual~$\widehat{\cC}$.
Indeed, it is not hard to see that $\widetilde{\cC}$ is the time reversal of the module-theoretic
dual~$\widehat{\cC}$, or, equivalently, $\widetilde{\cC}$ is the module-theoretic dual of the time reversal
of~$\cC$, \cite[Thm.~2.64]{JoZi99}.
Here, the time reversal code is obtained from the primary code by reversing the time axis.
In our purely polynomial setting, the reversal code can simply be defined as follows.
If $G\in\F[D]^{k\times n}$ is a minimal encoder of the code~$\cC$ with row degrees $\delta_1,\ldots,\delta_k$,
then it is easy to see that the {\em reciprocal matrix\/}
\begin{equation}\label{e-Gtilde}
   G':=\begin{pmatrix}D^{\delta_1}& & \\ &\ddots& \\ & &D^{\delta_k}\end{pmatrix}
                  G(D^{-1})\in\F[D]^{k\times n}
\end{equation}
is minimal and basic as well and has the same row degrees $\delta_1,\ldots,\delta_k$.
The code $\rev(\cC):=\im G'$ is called the {\em reversal code\/} of~$\cC$.
Thus, the above may be summarized as
\begin{equation}\label{e-Creversal}
   \widetilde{\cC}=\rev(\widehat{\cC}) = \widehat{\rev(\cC)}.
\end{equation}
We briefly wish to mention that in~\cite{Mi95} yet another notion of duality has been introduced, based on
local branch groups.
It is lengthy, but straightforward to show that for convolutional codes this type of duality is identical to
sequence space duality.
We omit the details, but only want to point out that the definition via local branch groups as given
in~\cite{Mi95} has the advantage to circumvent certain finiteness issues arising for sequence space duality
in the Laurent series setting.

In the rest of this section we will derive a MacWilliams Identity for the dual pairing~$(\cC,\,\widetilde{\cC})$.
Due to the close relationship between the module-theoretic dual and the sequence space dual this can indeed be
deduced from our previous result.
Since a time reversal of the state space system in\eqnref{e-SSS} essentially amounts to swapping~$X$
and~$Y$ in\eqnref{e-LambdaXY}, it should be intuitively clear that the WAM of~$\rev(\cC)$
will essentially be the transposed of the WAM of~$\cC$.
However, this is true only when choosing the right state space representations.
This will be carried out in the following computations.
In a first step we need a CCF of~$\rev(\cC)$.

\begin{proposition}\label{P-CCFreverse}
Let the data be as in Definition~\ref{D-CCF}.
Furthermore, let~$G'$ be the reciprocal matrix of~$G$ as in\eqnref{e-Gtilde}.
Then the CCF of~$G'$ is given by
$(A,B,C',E')$, where
\[
  \begin{pmatrix}C'\\ E'\end{pmatrix}
  =\begin{pmatrix} RA\T&RB\T\\BR&I-BB\T\end{pmatrix}\begin{pmatrix}C\\E\end{pmatrix}
\]
and
\begin{equation}\label{e-Rmatrix}
   R=\begin{pmatrix}R_1 & & \\[-1ex] &\ddots& \\[-1ex] & & R_{r}\end{pmatrix}\in\mbox{\rm{GL}}_{\delta}(\F) \text{ with }
   R_i=\begin{pmatrix} & &\!\! 1\\[-.7ex] &\quad \cdot&\\[-2ex] &\cdot& \\[-2ex] &\cdot\quad&\\[-1ex] 1\!\!& & \end{pmatrix}
   \in\mbox{\rm{GL}}_{\delta_i}(\F).
\end{equation}
Moreover, the matrix $L:=\Smallfourmat{RA\T}{RB\T}{BR}{I-BB\T}$ satisfies $LL\T=I_{\delta+k}$,
thus $L\in\mbox{\rm{GL}}_{\delta+k}(\F)$.
\end{proposition}
\begin{proof}
Since~$G$ and $G'$ are both minimal with the same row degrees
$\delta_1,\ldots,\delta_{k}$, it is clear that the CCF of~$G'$
has, just like~$G$, state transition matrix~$A$ and input-to-state matrix~$B$.
The identities for~$C'$ and~$E'$ follow straightforwardly from the form of~$A,\,B,\,C,\,E$
as in Definition~\ref{D-CCF} along with the simple matrix identities
\begin{equation}\label{e-RA}
 A\T A+B\T B=I,\quad I-BB\T=\Smallfourmat{0}{0}{0}{I_{k-r}},\quad R=R^{-1}=R\T,\text{ and } RA\T R=A
\end{equation}
as well as the fact that
\[
  C'=\begin{pmatrix}C'_1\\[1ex] \vdots \\[1.5ex] C'_r\end{pmatrix}, \text{ where }
  C'_i=\begin{pmatrix}g_{i,\delta_i-1}\\ \vdots\\ g_{i,1}\\ g_{i,0}\end{pmatrix},
  \text{ and }
  E'=\begin{pmatrix} g_{1,\delta_1}\\ \vdots\\ g_{r,\delta_r}\\ g_{r+1,0}\\ \vdots \\ g_{k,0}\end{pmatrix}.
\]
The identity $LL\T=I_{\delta+k}$ can easily be verified using\eqnref{e-RA} and
Remark~\ref{R-abceprop}(i).
\end{proof}

Now it is easy to present the WAM of the reversal code.
\begin{corollary}\label{C-Lambda'}
Let~$\Lambda$ be the WAM of~$\cC$ associated with the CCF $(A,B,C,E)$.
Then the WAM~$\Lambda'$ of the reversal code~$\rev(\cC)$ associated with the CCF
$(A,B,C',E')$ given in Proposition~\ref{P-CCFreverse} satisfies
\begin{equation}\label{e-LambdaLambda'}
  \Lambda'_{X,Y}=\Lambda_{YR,XR}\text{ for all }(X,\,Y)\in\cF,
\end{equation}
where~$R\in\mbox{\rm{GL}}_{\delta}(\F)$ is as in\eqnref{e-Rmatrix}.
\end{corollary}
\begin{proof}
From\eqnref{e-DeltaLambda} we have
$\Lambda'_{X,Y}\not=0\Longleftrightarrow(X,Y)\in\Delta$ and
$\Lambda_{YR,XR}\not=0\Longleftrightarrow (YR,XR)\in\Delta$.
Hence we first have to show that $(X,Y)\in\Delta\Longleftrightarrow(YR,XR)\in\Delta$.
Using the definition of~$\Delta$ in\eqnref{e-DeltaOmega} as well as the matrix~$L$ from
Proposition~\ref{P-CCFreverse} and the identities in\eqnref{e-RA} we obtain
\[
  \im\begin{pmatrix}I&A\\0&B\end{pmatrix}\begin{pmatrix}0&R\\R&0\end{pmatrix}
  =\im\begin{pmatrix}AR&R\\BR&0\end{pmatrix}=\im L^{-1}\begin{pmatrix}I&A\\0&B\end{pmatrix}
  =\im\begin{pmatrix}I&A\\0&B\end{pmatrix}.
\]
This shows that $(X,Y)\in\Delta$ iff $(YR,XR)\in\Delta$ and
it remains to prove\eqnref{e-LambdaLambda'} for $(X,Y)\in\Delta$.
From \cite[Lem.~III.8]{GS08} we know that for $(X,Y)\in\Delta$
\begin{equation}\label{e-LLambdaXY}
  \Lambda'_{X,Y}=\we(XC'+YB\T E'+\cC'_{\text{\rm const}})\text{ and }
  \Lambda_{YR,XR}=\we\big((YR)C+(XR)B\T E+\cCconst\big).
\end{equation}
Since $\cCconst$ is generated by the constant rows of the minimal encoder~$G$, it follows directly from the
definition of the reciprocal matrix in\eqnref{e-Gtilde} that $\cC'_{\text{\rm const}}=\cCconst$.
Furthermore, since $(X,Y)\in\Delta$, there exists $u\in\F^k$ such that $Y=XA+uB$.
Using Proposition~\ref{P-CCFreverse} and\eqnref{e-RA} as well as Remark~\ref{R-abceprop}(i) we compute
\begin{align*}
  XC'+YB\T E'&=XRA\T C+XRB\T E+YB\T BRC+YB\T(I-BB\T)E=\\
            &=XARC+XRB\T E+XAB\T BRC+uBB\T BRC\\
            &=(XA+uB)RC+(XR)B\T E=(YR)C+(XR)B\T E
\end{align*}
With the aid of\eqnref{e-LLambdaXY} this proves\eqnref{e-LambdaLambda'} for all $(X,\,Y)\in\cF$.
\end{proof}

Now it is straightforward to formulate and prove a MacWilliams Identity for the sequence space dual code.
The transformation matrix~$Q$, needed for the identity, will be given explicitly in the proof.

\begin{theorem}\label{T-MacWsequence}
Let~$\cC\subseteq\F[D]^n$ be as in General Assumption~\ref{A-data1} and let $\widetilde{\cC}$ be the sequence
space dual of~$\cC$.
Let $(\tilde{A},\tilde{B},\tilde{C},\tilde{E})$ be a CCF of~$\widetilde{\cC}$ and let
$\widetilde{\Lambda}$ be the associated WAM.
Then there exists a matrix $Q\in\mbox{\rm{GL}}_{\delta}(\F)$ such that
\[
   \widetilde{\Lambda}_{XQ,YQ}=q^{-k}\HH(\cH\Lambda\cH^{-1})_{X,Y}\text{ for all }X,Y\in\F^{\delta}.
\]
As a consequence, the generalized WAM's of~$\cC$ and $\widetilde{\cC}$ satisfy
$[\widetilde{\Lambda}]=q^{-k}\HH(\cH[\Lambda]\cH^{-1})$.
\end{theorem}
\begin{proof}
Let~$\cC'=\rev(\cC)$ and let~$\Lambda'$ be the WAM associated with the
CCF $(A,B,C',E')$, where $C',\,E'$ are as in Proposition~\ref{P-CCFreverse}.
Then $\widetilde{\cC}=\widehat{\cC'}$, due to\eqnref{e-Creversal}.
Recall the matrix~$R$ from\eqnref{e-Rmatrix}.
By Theorem~\ref{T-MacWID}' we have
\[
  \widetilde{\Lambda}_{XR\tilde{P}^{-1},YR\tilde{P}^{-1}}=q^{-k}\HH(\cH\Lambda'\T\cH^{-1})_{XR,YR}
  \text{ for all }(X,Y)\in\cF,
\]
where $\tilde{P}:=\tilde{C}E'\T B-\tilde{N}A$ and $\tilde{N}$ is as in\eqnref{e-N} with $S_j$ and $\widehat{S}_j$
replaced by $S'_j$ and $\tilde{S}_j$ defined via the CCF's $(A,B,C',E')$ and $(\tilde{A},\tilde{B},\tilde{C},\tilde{E})$,
respectively, as in\eqnref{e-S}.
Let $\cP:=\cP(R)$, see\eqnref{e-PP}.
Using\eqnref{e-Pitrafo} we obtain
\[
    (\cH\Lambda'\T\cH^{-1})_{XR,YR}=\Big((\cP\cH\cP^{-1})(\cP\Lambda'\T\cP^{-1})(\cP\cH\cP^{-1})^{-1}\Big)_{X,Y}.
\]
But now the definition of~$\cH$ in\eqnref{e-cH} along with\eqnref{e-Pitrafo} and the identity $RR^{\sf T}=I$ shows that
$\cP\cH\cP^{-1}=\cH$.
Moreover, by Corollary~\ref{C-Lambda'} and\eqnref{e-Pitrafo} we have $\cP\Lambda'\T\cP^{-1}=\Lambda$.
All this establishes Theorem~\ref{T-MacWsequence} with the transformation matrix $Q:=R\tilde{P}^{-1}$.
\end{proof}

\section*{Conclusion}
We established a MacWilliams Identity for convolutional codes.
It consists of a conjugation of the weight adjacency matrix followed by the entrywise
MacWilliams transformation for block codes.
The identity applies to both module-theoretic duality as well as the sequence space duality.
The result opens the door to investigating self-dual convolutional codes (with respect to any
duality notion) with the aid of invariant theory.
This will be pursued in a future project.

\appendix
\section*{Appendix A}
\setcounter{section}{1}
\setcounter{theorem}{0}
\renewcommand{\theequation}{A.\arabic{equation}}
\setcounter{equation}{0}
In this section we prove the purely matrix theoretical results of Proposition~\ref{P-abce}.
As before, the data are as in General Assumption~\ref{A-data1}.
Since $G=E+\sum_{i\geq1}BA^{i-1}CD^i$
we have $B^{\sf T}G=\sum_{i\geq0}S_iD^i$ with the matrices~$S_i$ as given in\eqnref{e-S}.
Thus, $\widehat{B}^{\sf T}\widehat{G}G^{\sf T}B=0$ implies
\begin{equation}\label{e-SShat}
  \sum_{i=0}^m\widehat{S}_iS_{m-i}^{\sf T}=0\text{ for all }m\geq0.
\end{equation}
Using the CCF, it is easy to see that
$\sum_{i\geq 1}(BA^{i-1})^{\sf T}(BA^{i-1})=I_{\delta}$.
This in turn yields
\begin{equation}\label{e-C}
  C=\sum_{i\geq 1}(A^{\sf T})^{i-1}S_i
\end{equation}
and, consequently,
\begin{equation}\label{e-CChat}
  \widehat{C}C^{\sf T}=\sum_{m\geq2}\sum_{i=1}^{m-1}
           (\widehat{A}^{\sf T})^{i-1}\widehat{S}_iS_{m-i}^{\sf T}A^{m-i-1}.
\end{equation}
Now we are ready for the

\noindent{\sc Proof of Proposition~\ref{P-abce}:}
(a) From\eqnref{e-SShat} we obtain
\[
  \widehat S_iS_{m-i}^{\sf T}=-\sum_{j=0}^{i-1}\widehat S_jS_{m-j}^{\sf T}
    -\sum_{j=i+1}^m \widehat S_j S_{m-j}^{\sf T}
\]
for $i=1,\ldots,m-1$.
Using\eqnref{e-CChat} and~$N$ and~$\widehat{N}^{\sf T}$ from\eqnref{e-N} and\eqnref{e-Nhattrans}
we therefore compute
\begin{align*}
-\widehat CC^{\sf T}&=-\sum_{m\geq 2}\sum_{i=1}^{m-1}
                        (\widehat A^{\sf T})^{i-1}\widehat S_iS_{m-i}^{\sf T}A^{m-(i+1)}\\
              &=\sum_{m\geq 2}\sum_{i=1}^{m-1}(\widehat A^{\sf T})^{i-1}
                \Big(\sum_{j=0}^{i-1}\widehat S_jS_{m-j}^{\sf T}+\sum_{j=i+1}^m\widehat S_jS_{m-j}^{\sf T}\Big)
                 A^{m-(i+1)}=N+\widehat N^{\sf T},
\end{align*}
which is what we wanted.
\\
(b) Using again\eqnref{e-Nhattrans} one obtains
\begin{align*}
  NA+\widehat A^{\sf T}\widehat N^{\sf T}
      &=\sum_{m\geq 2}\bigg(
          \sum_{i=1}^{m-1}\sum_{j=0}^{i-1} (\widehat A^{\sf T})^{i-1}\widehat S_jS_{m-j}^{\sf T}A^{m-i}
         +\sum_{i=1}^{m-1}\sum_{j=i+1}^m(\widehat A^{\sf T})^{i}\widehat S_jS_{m-j}^{\sf T}A^{m-(i+1)}\bigg) \\
      &=\sum_{m\geq 2}\bigg(
          \sum_{i=0}^{m-2}\sum_{j=0}^i(\widehat A^{\sf T})^{i}\widehat S_jS_{m-j}^{\sf T}A^{m-(i+1)}
         +\sum_{i=1}^{m-1}\sum_{j=i+1}^m(\widehat A^{\sf T})^{i}\widehat S_jS_{m-j}^{\sf T}A^{m-(i+1)}\bigg)\\
     &=\sum_{m\geq 2}\bigg(
           \sum_{i=1}^{m-2}(\widehat A^{\sf T})^i\Big(\sum_{j=0}^{m}\widehat S_jS_{m-j}^{\sf T}\Big)A^{m-(i+1)}
          +\widehat S_0S_m^{\sf T}A^{m-1}+(\widehat A^{\sf T})^{m-1}\widehat S_mS_0^{\sf T}\bigg)
\end{align*}
Due to\eqnref{e-SShat} the inner sum over~$j$ vanishes, and adding
$0=\widehat S_0S_1^{\sf T}+\widehat S_1S_0^{\sf T}$, which is\eqnref{e-SShat} for $m=1$, we proceed with
\begin{align*}
 NA+\widehat A^{\sf T}\widehat N^{\sf T}
     &= \sum_{m\geq 2}
         \left(\widehat S_0S_m^{\sf T}A^{m-1}+(\widehat A^{\sf T})^{m-1}\widehat S_mS_0^{\sf T}\right)
         +\widehat S_0S_1^{\sf T}+\widehat S_1S_0^{\sf T}\\
     &=\sum_{m\geq 1}
        \left(\widehat S_0S_m^{\sf T}A^{m-1}+(\widehat A^{\sf T})^{m-1}\widehat S_mS_0^{\sf T}\right)
      =\widehat S_0C^{\sf T}+\widehat CS_0^{\sf T},
\end{align*}
where the last identity is a consequence of\eqnref{e-C}.
This proves part~(b).
\\
(c) As before let $e_1,\ldots,e_{\delta}$ be the standard basis vectors of~$\F^{\delta}$.
Throughout the rest of this proof denote, for any matrix~$M$, the $\gamma$-th column (resp.\ $\gamma$-th row)
of~$M$ by $M_{(\gamma)}$ (resp.\ $M^{(\gamma)}$).
Let us assume that the matrices~$A$ and~$B$ are as in Definition~\ref{D-CCF}.
Then we have $\ker A=\text{span}_{\F}\{e_{j_l}\mid l=1,\ldots,r\}$, where
$j_l=\sum_{a=1}^{l}\delta_a$. Moreover, $AA^{\sf T}$ is the diagonal matrix with $(AA^{\sf T})_{j_l,j_l}=0$
for $l=1,\ldots,r$ and $(AA^{\sf T})_{i,i}=1$ else.
Therefore, it suffices to show that the $\mu$-th column of $N$ is zero for all
$\mu\in\{j_1,\ldots,j_r\}$.
Thus, let $\mu=\sum_{a=1}^l\delta_a$ for some $l=1,\ldots,r$.
In order to prove the desired result we will even show that
\begin{equation}\label{e-column}
  (S^{\sf T}_{m-j}A^{m-i-1})_{(\mu)}=0\text{ for all $m\geq2$ and $1\leq i\leq m-1$ as well as $0\leq j\leq i-1$.}
\end{equation}
This, of course, implies $N_{(\mu)}=0$ due to\eqnref{e-N}.
In order prove\eqnref{e-column} notice that
\[
   S^{\sf T}_{m-j}A^{m-i-1}=C^{\sf T}(A^{\sf T})^{m-j-1}B^{\sf T}BA^{m-i-1}.
\]
Put $\nu:=\sum_{a=1}^{l-1}\delta_a+1$.
The definition of $A$ and $B$ shows that
\[
  (BA^{m-i-1})_{(\mu)}\neq 0\Longleftrightarrow (A^{m-i-1})_{\nu,\mu}=1
  \Longleftrightarrow \delta_l-1=m-i-1\Longleftrightarrow i=m-\delta_l.
\]
Hence for $i\not=m-\delta_l$ we have $(BA^{m-i-1})_{(\mu)}=0$ and it remains to prove\eqnref{e-column} for the case
$i=m-\delta_l$.
In that case $0\leq j\leq m-\delta_l-1$ implies $\delta_l<m-j$.
Using that $(B^{\sf T}B)^{(\nu)}=e_{\nu}$, the $\nu$-th row of $S_{m-j}$ is
\[
  (S_{m-j})^{(\nu)}=(B^{\sf T}BA^{m-j-1}C)^{(\nu)}=(A^{m-j-1}C)^{(\nu)}=(A^{m-j-1})^{(\nu)}C=0,
\]
where the last identity follows from the simple fact that the $l$-th diagonal block of
$A^{m-j-1}$ is zero, as $m-j-1\geq \delta_l$.
Transposing the obtained identity yields $(S_{m-j}^{\sf T})_{(\nu)}=0$.
Since $(A^{m-i-1})_{(\mu)}=(A^{\delta_l-1})_{(\mu)}=e_{\nu}^{\sf T}$, we obtain
\[
   (S_{m-j}^{\sf T}A^{m-i-1})_{(\mu)}=S_{m-j}^{\sf T}(A^{m-i-1})_{(\mu)}=(S_{m-j}^{\sf T})_{(\nu)}=0.
\]
This proves\eqnref{e-column} for the case $i=m-\delta_l$ and thus concludes the proof of
Proposition~\ref{P-abce}.
\hfill$\Box$

\bibliographystyle{abbrv}
\bibliography{literatureAK,literatureLZ}

\begin{thebibliography}{10}

\bibitem{Ab92}
K.~A.~S. Abdel-Ghaffar.
\newblock On unit constrained-length convolutional codes.
\newblock {\em IEEE Trans. Inform. Theory}, IT-38:200--206, 1992.

\bibitem{Fo73}
{G.~D. Forney, Jr.}
\newblock Structural analysis of convolutional codes via dual codes.
\newblock {\em IEEE Trans. Inform. Theory}, IT-19:512--518, 1973.

\bibitem{Fo75}
{G.~D. Forney, Jr.}
\newblock Minimal bases of rational vector spaces, with applications to
  multivariable linear systems.
\newblock {\em SIAM J. on Contr.}, 13:493--520, 1975.

\bibitem{Fo91}
{G.~D. Forney, Jr.}
\newblock Algebraic structure of convolutional codes, and algebraic system
  theory.
\newblock In {\em Mathematical System Theory, The influence of R.~E. Kalman
  {\rm (A. Antoulas, ed.)}}, pages 527--557. Springer, 1991.

\bibitem{Fo01}
{G.~D. Forney, Jr.}
\newblock Codes on graphs: {N}ormal realizations.
\newblock {\em IEEE Trans. Inform. Theory}, IT-47:520--548, 2001.

\bibitem{FT04}
{G.~D. Forney, Jr.} and M.~D. Trott.
\newblock The dynamics of group codes: {D}ual abelian group codes and systems.
\newblock {\em IEEE Trans. Inform. Theory}, IT-50:2935--2965, 2004.

\bibitem{GL05p}
H.~Gluesing-Luerssen.
\newblock On the weight distribution of convolutional codes.
\newblock {\em Linear Algebra and its Applications}, 408:298--326, 2005.

\bibitem{GS08}
H.~Gluesing-Luerssen and G.~Schneider.
\newblock On the {M}ac{W}illiams identity for convolutional codes.
\newblock {\em IEEE Trans. Inform. Theory}, IT-54:1536--1550, 2008.

\bibitem{HJZ02}
S.~H{\"o}st, R.~Johannesson, and V.~V. Zyablov.
\newblock Woven convolutional codes {I}: {E}ncoder properties.
\newblock {\em IEEE Trans. Inform. Theory}, IT-48:149--161, 2002.

\bibitem{HP03}
W.~C. Huffman and V.~Pless.
\newblock {\em Fundamentals of Error-Correcting Codes}.
\newblock Cambridge University Press, Cambridge, 2003.

\bibitem{JSW00}
R.~Johannesson, P.~St{\aa}hl, and E.~Wittenmark.
\newblock A note on type {II} convolutional codes.
\newblock {\em IEEE Trans. Inform. Theory}, IT-46:1510--1514, 2000.

\bibitem{JoZi99}
R.~Johannesson and K.~S. Zigangirov.
\newblock {\em Fundamentals of Convolutional Coding}.
\newblock IEEE Press, New York, 1999.

\bibitem{JPB90}
J.~Justesen, E.~Paaske, and M.~Ballan.
\newblock Quasi-cyclic unit memory convolutional codes.
\newblock {\em IEEE Trans. Inform. Theory}, IT-36:540--547, 1990.

\bibitem{LC83}
S.~Lin and D.~J. Costello~Jr.
\newblock {\em Error Control Coding: {F}undamentals and Applications}.
\newblock Prentice Hall, 1983.

\bibitem{MS77}
F.~J. Mac{W}illiams and N.~J.~A. Sloane.
\newblock {\em The Theory of Error-Correcting Codes}.
\newblock North-Holland, 1977.

\bibitem{MaSa67}
J.~L. Massey and M.~K. Sain.
\newblock Codes, automata, and continuous systems: {E}xplicit interconnections.
\newblock {\em IEEE Trans. Aut. Contr.}, AC-12:644--650, 1967.

\bibitem{McE98}
R.~J. Mc{E}liece.
\newblock The algebraic theory of convolutional codes.
\newblock In V.~S. Pless and W.~C. Huffman, editors, {\em Handbook of Coding
  Theory, Vol.~1}, pages 1065--1138. Elsevier, Amsterdam, 1998.

\bibitem{McE98a}
R.~J. Mc{E}liece.
\newblock How to compute weight enumerators for convolutional codes.
\newblock In M.~Darnell and B.~Honory, editors, {\em Communications and Coding
  (P.~G.~Farrell 60th birthday celebration)}, pages 121--141. Wiley, New York,
  1998.

\bibitem{Mi95}
T.~Mittelholzer.
\newblock Convolutional codes over groups: {A} pragmatic approach.
\newblock In {\em Proc. of the 33rd Allerton Conference on Communications,
  Control, and Computing}, pages 380--381, 1995.

\bibitem{SM77}
J.~B. Shearer and R.~J. Mc{E}liece.
\newblock There is no {M}ac{W}illiams identity for convolutional codes.
\newblock {\em IEEE Trans. Inform. Theory}, IT-23:775--776, 1977.

\bibitem{Vi71}
A.~J. Viterbi.
\newblock Convolutional codes and their performance in communication systems.
\newblock {\em IEEE Trans. Commun. Technol.}, COM-19:751--772, 1971.

\end{thebibliography}
\end{document}